\documentclass[twocolumn,amsmath,amssymb,showpacs]{revtex4}
\usepackage{graphicx}

\usepackage{xcolor}
\usepackage{longtable}
\usepackage{times}
\usepackage{dcolumn}
\usepackage{bm}
\usepackage[plainpages = false, pdfpagelabels, 
                 bookmarks,
                 bookmarksopen = true,
                 bookmarksnumbered = true,
                 linktocpage,
                 pagebackref=false,
                 colorlinks = true,
                 linkcolor = blue,
                 urlcolor  = blue,
                 citecolor = blue,
                 anchorcolor = green,
                 hyperindex = true,
                 hyperfigures
                 ]{hyperref} 
\begin{document}

\title {\Large Theoretical investigation of charge-changing cross section and interaction cross section
	for Be, B, C, N, O, and F isotopes on $^{12}$C at 200\textendash-1050 MeV/nucleon}

\author{Z. \surname{Hasan}$^1$}\email{zafaramu@gmail.com}
\author{M. \surname{Imran}$^{2,3}$}\email{imran.phys.amu@gmail.com}
\author{A. A. \surname{Usmani}$^3$}\email{anisul@iucaa.ernet.in}
\author{Z. A. \surname{Khan}$^3$}\email{zakhan.amu@gmail.com}
\affiliation{$^1$Department of Applied Physics, ZH College of Engineering and Technology, Aligarh Muslim University, Aligarh-202002, India}
\affiliation{$^2$Applied Science \& Humanities Section, University Women's Polytechnic, Aligarh Muslim University, Aligarh-202002, India}
\affiliation{$^3$Department of Physics, Aligarh Muslim University, Aligarh-202002, India}

\begin{abstract}

To establish credibility for the use of the Slater determinant harmonic oscillator (SDHO) density in predicting
root-mean-square proton and neutron radii for Be, B, C, N, O, and F isotopes [M. Imran et al., {\color{blue}Phys. Rev.
C {\bf 110}, 014623 (2024)}], in this work we propose to study charge-changing and interaction cross sections for
the said isotopes on  $^{12}$C at a wider range of incident energies (200-1050 MeV/nucleon), involving different
density distributions; the calculations also assess the importance of nuclear medium effects. Working within
the framework of the Glauber model, we involve two-parameter Fermi (2pF) and three-parameter Fermi (3pF)
shapes of density distributions, and use the in-medium as well as free behavior of the nucleon-nucleon (NN)
amplitude. The results provide enough basis to support the matter radii of exotic isotopes obtained using SDHO
densities.

\end{abstract}
\maketitle

\section{Introduction}
\label{sec1}

The reliable estimates for root-mean-square (rms) proton and neutron radii of short-lived nuclei away from the stability line is a challenging task in the study of nuclear reactions. The conventional tools for measuring the proton radii, the electron scattering and isotope-shift measurements, though have been quite useful for probing a large section of stable nuclei, but they are limited to only few unstable nuclei. Due to this, the study of some of the important aspects of unstable (short-lived) nuclei (e.g., the existence of thick neutron skins and halos \cite{1,2,3,4}) demands to involve other sources of measurements to provide matter radii of such nuclei. Both, thick neutron skin and nuclear halo in exotic neutron-rich nuclei exhibit large difference between neutron and proton radii as compared to their neighboring nuclei, thereby involve a large spatial separation of one or two weakly bound valence neutrons and hence form a low-density region around the core nucleus. It is, thus, important to have equally reliable values of proton and neutron radii in order to have proper understanding of thick neutron skin and neutron halo in exotic nuclei.

Earlier and also in recent past \cite{4,5,6,7,8,9,10,11}, the measurements of the charge-changing cross sections (CCCSs) were performed with a view that these experiments may be used to provide information about the proton radii of unstable (exotic) nuclei. Further, it was hoped that the combination of the study of CCCSs and reaction (interaction) cross sections could be a better choice to have reliable predictions for proton and neutron radii of exotic nuclei.
 
On theoretical front, it has been observed that the Glauber model has been quite successful in extracting the matter radii of exotic nuclei \cite{12}, utilizing the experimental data on interaction cross sections. However, contrary to our belief that CCCSs may directly provide information about the proton radii of unstable nuclei, calculations of CCCSs using the Glauber model \cite{6,13,14,15,16,17} suggest that the presence of neutrons can not be ignored in explaining the CCCS data, making difficult to formulate the reaction mechanism for CCCS. Keeping this in view, an alternative way of calculations has been suggested, according to which the study of CCCSs could be possible in terms of a phenomenological scaling (correction) parameter \cite{6,13,18} that accommodates the contribution due to presence of neutrons in the projectile. The idea of scaling parameter in CCCS calculations has also been considered, but in a different way, in our recent work \cite{19}, and the Glauber model calculations have been carried out to extract root-mean-square proton and neutron radii for Be, B, C, N, O, and F isotopes from the analysis of their charge-changing and interaction cross sections. The extracted proton and neutron radii \cite{19} have been examined in view of some important features such as the neutron skin thickness, halo-like structure, and subshell closure observed in neutron-rich unstable nuclei. Our results \cite{19} are found to agree fairly well with those available in literature.

In this work, we proposed to verify and improve utility of our recent Glauber model findings on rms proton, neutron, and matter radii for Be, B, C, N, O, and F isotopes \cite{19} obtained from the analysis of charge-changing and interaction cross sections involving Slater determinant harmonic oscillator (SDHO) density and in-medium behavior of the nucleon-nucleon (NN) amplitude. For this purpose, we have now considered two-parameter Fermi (2pF) and three-parameter Fermi (3pF) shapes for the density distributions and perform the Glauber model calculations for charge-changing and interaction cross section at a wider range of incident energies (200-1050 MeV/nucleon), involving the in-medium as well as free behavior of the NN amplitude; the calculations also assess the importance of nuclear medium effects. Our aim is to see how far the present work provides support to using SDHO density in predicting reliable estimates for proton and neutron radii of exotic nuclei.  
                        
The formulation of the problem is given in  Sec. \ref{sec2}. The numerical results are presented and discussed in Sec. \ref{sec3}. The conclusions are 
presented in Sec. \ref{sec4}.

\section{Formulation}

\label{sec2}
\subsection{Reaction (interaction) cross section}

According to the Glauber model, the reaction cross section ($\sigma_{R}$) for nucleus-nucleus collision is given by
\begin{equation}
\sigma_{R}= 2\pi\int \left[1- \vert S_{el}(b)\vert^{2}\right]b~db,
\label{eq1}
\end{equation}
%\begin{widetext}
\begin{equation}
S_{el}(b)= \langle \psi_{T}\psi_{P}\vert\prod^{A}_{i=1}\prod^{B}_{j=1}[1-\Gamma_{NN}(\vec{b}-\vec{s_{i}}+\vec{s^{'}_{j}})]\vert\psi_{P}\psi_{T}\rangle,
\label{eq2}
\end{equation}
%\end{widetext}
where  $\psi_{P}$ ($\psi_{T}$) is the ground-state wave function of projectile (target) nucleus, $A(B)$ is the mass number of target (projectile) nucleus, $\vec{b}$ is the impact parameter vector perpendicular to the incident momentum, $\vec {s_{i}}\vec{(s^{'}_{j}})$ are the projections of
target (projectile) nucleon coordinates on the impact parameter plane, 
and $\Gamma_{NN}({b})$ is the $NN$ profile function, which is related to the $NN$ scattering amplitude $f_{NN}(q)$ as 
follows  
\begin{equation}
\Gamma_{NN}({b})= \frac{1}{2\pi ik}~\int \exp(-i\vec {q}.\vec {b})f_{NN}({q})~d^{2}q, 
\label{eq3}
\end{equation}
where $k$ is the incident nucleon momentum corresponding to the projectile kinetic energy per nucleon, and $\vec{q}$ is the momentum transfer.

Following the approach of Ahmad \cite{20}, the S-matrix element, $S_{el}({b})$, up to two-body correlation (density) term takes the following form:
\begin{equation}
S_{el}({b})\approx S_{0}(b)+S_{2}(b),
\label {eq4}
\end{equation}
where
\begin{equation}
S_{0}({b})= [1-\Gamma^{NN}_{00}(b)]^{AB}, 
\label{eq5}
\end{equation}
and
\begin{gather}
%\begin{widetext}
%\begin{equation}
S_{2}({b})=\Bigl\langle\psi_{T}\psi_{P}\Bigl|\frac{1}{2!}[1-\Gamma^{NN}_{00}(b)]^{AB-2}~~~~~~~~~~~~~~~~~~\nonumber\\
~~~~~~~~~~~~~~~~~~\times \sum^{'}_{i_{1},j_{1}}\sum^{'}_{i_{2},j_{2}}\gamma_{i_{1},j_{1}}\gamma_{i_{2},j_{2}}\Bigr|\psi_{P}\psi_{T}\Bigr\rangle,
\label{eq6}
%\end{equation}
%\end{widetext} 
\end{gather}
with 
\begin{equation}
\gamma_{ij}=\Gamma^{NN}_{00}(\vec{b})-\Gamma_{NN}(\vec{b}-\vec{s_{i}}+\vec{s^{'}_{j}}), 
\label{eq7}
\end{equation}
and
\begin{equation}
\Gamma^{NN}_{00}({b})=\int \rho_{T}(\vec r)~\rho_{P}(\vec{r^{'}})~\Gamma_{NN}(\vec b-\vec s +\vec{s^{'}})~d\vec r~d\vec{r^{'}}.
\label {eq8}
\end{equation}
The primes on the summation signs in Eq. (\ref{eq6}) indicate the restriction that two pairs of indices can not be equal at the same time (for example,
if $i_{1}$ = $i_{2}$ then $j_{1} \neq j_{2}$ and vice versa). The quantities $\rho_{T}$ and $\rho_{P}$ in Eq. (\ref{eq8}) are the (one-body) ground state
densities of the target and projectile, respectively. From calculations point of view, it should be noted that the distinction between protons and neutrons in both the projectile and target has been included in the uncorrelated part ($S_{0}$) of the S-matrix element only. This makes the use of different values of the parameters for pp and pn scattering amplitudes and considers different density distributions for protons and neutrons in the colliding nuclei. However, it turns out that inclusion of the distinct features of protons and neutrons in the two-body correlation term ($S_{2}({b})$) is not as straightforward as it was in $S_{0}({b})$. Therefore, the two-body correlation term uses the average behavior of pp and pn interactions, and involves matter density distributions. More explicitly, the evaluation of $S_{0}({b})$ and $S_{2}({b})$ leads to the following expressions:
\begin{gather}
	S_{0}({b})= [1-\Gamma^{pp}_{00}({b})]^{Z_{P}Z_{T}}[1-\Gamma^{np}_{00}({b})]^{N_{P}Z_{T}}~~~\nonumber\\ 
	~~~~~~~~~~~~~~~~~\times[1-\Gamma^{pn}_{00}({b})]^{Z_{P}N_{T}}[1-\Gamma^{nn}_{00}({b})]^{N_{P}N_{T}},
	%\end{multline*}
	\label{eq9}
\end{gather}
\begin{gather}
	S_{2}({b})=-\frac{AB}{8\pi^{2}k^{2}}[1-\Gamma^{NN}_{00}(b)]^{AB-2}[(A-1)(B-1)\nonumber\\
	~~~~~~~~~~~~~~~~~\times(G_{22}(b)-G_{00}^{2}(b))+(B-1)\nonumber\\
	~~~~~~~~~~~~~~~~~\times(G_{21}(b)-G_{00}^{2}(b))+(A-1)\nonumber\\
	~~~~~~\times(G_{12}(b)-G_{00}^{2}(b))],
	\label{eq10}
	%\end{equation}
	%\end{widetext}
	%\end{multline*}
\end{gather}
with
\begin{equation}
\Gamma^{ij}_{00}({b})=\int \rho^{j}_{T}(\vec r_{j})~\rho^{i}_{P}(\vec{r_{i}^{'}})~\Gamma_{ij}(\vec b-\vec s_{j} +\vec{s_{i}^{'}})~d\vec r_{j}~d\vec{r_{i}^{'}},
\label{eq11}
\end{equation}
where $Z_{T}(Z_{P})$ and $N_{T}(N_{P})$ are the target(projectile) atomic and neutron number, respectively, and each of $i$ and $j$ stands for a proton and a neutron. For details of $G_{22}(b), G_{21}(b), G_{12}(b)$, and  $G_{00}(b)$, we refer readers to follow the work presented in Ref. \cite{19}.  

\subsection{Charge-changing cross section}

In addition to the total reaction cross section ($\sigma_{R}$), the charge-changing cross section ($\sigma_{cc}$) and the total neutron removal cross section ($\sigma_{-xn}$) are another important observables. The reaction cross section is a measure of the net probability of particle removal from the projectile; the charge-changing cross section is defined as the total probability for change in the charge number of the projectile, while the neutron removal cross section can be similarly defined as the cross section for the processes which result in a change of the neutron number of the projectile. Thus, the total reaction cross section $\sigma_{R}$ is the sum of $\sigma_{cc}$ and $\sigma_{-xn}$.

As pointed out earlier, unlike the reaction cross section, the Glauber model calculations for $\sigma_{cc}$ \cite{6,13,14,15,16,17} suggest that it is not easy to understand the reaction mechanism for charge-changing cross section due to partial involvement of the projectile neutrons. This status of $\sigma_{cc}$ has further suggested \cite{6,13,18} that the calculations of $\sigma_{cc}$ may be possible within the framework of Glauber model, provided one introduces a phenomenological scaling (correction) parameter that takes care of the presence of neutrons in the projectile. Such a consideration allows us to write $\sigma_{cc}$ as follows:   
\begin{equation}
\sigma_{cc}=\epsilon ~ \sigma^{p}_{cc},
\label{eq30}
\end{equation} 
where $\epsilon$ is the correction parameter which is defined as the ratio of the experimental $\sigma_{cc}$ and calculated $\sigma^{p}_{cc}$ values ($\sigma^{exp}_{cc}/\sigma^{p}_{cc}$). The quantity $\sigma^{p}_{cc}$ is the contribution to the charge-changing cross section due to the scattering of only projectile protons. Keeping in view the limitations of finding the correction parameter $\epsilon$ in the works of Bhagwat and Gambhir \cite{13},  Yamaguchi \textit{et al.} \cite{6} and Li \textit{et al.} \cite{18}, we have adopted \cite{19} a different prescription to get the correction parameter which will be used in the present work also. Thus, we need to mention only the calculation of $\sigma^{p}_{cc}$ in order to get the results for $\sigma_{cc}$. Following Eq. (\ref{eq1}), $\sigma^{p}_{cc}$ is given by     
\begin{equation}
\sigma^{p}_{cc}=2\pi\int \left[1- \vert S^{p}_{el}(b)\vert^{2}\right]b~db,
\label{eq31} 
\end{equation}
where
\begin{equation}
S^{p}_{el}({b})\approx S^{p}_{0}(b)+S^{p}_{2}(b).
\label{eq32}
\end{equation}
The quantities $S^{p}_{0}(b)$ and $S^{p}_{2}(b)$ that consider only the projectile protons can be obtained by putting $N_{P}$ = 0 in the respective expressions for $S_{0}(b)$[Eq. (\ref{eq9})] and $S_{2}(b)$[Eq. (\ref{eq10})]. This simplification leads to the following expressions for $S^{p}_{0}(b)$ and $S^{p}_{2}(b)$:
\begin{equation}
S^{p}_{0}(b)= [1-\Gamma^{pp}_{00}(b)]^{Z_{P}Z_{T}}[1-\Gamma^{pn}_{00}(b)]^{Z_{P}N_{T}}, 
\label{eq33}
\end{equation}
%\begin{widetext} 
%\begin{equation}
\begin{gather}
S^{p}_{2}(b)=-\frac{AZ_{P}}{8\pi^{2}k^{2}}(1-\Gamma^{NN}_{00}(b))^{AZ_{P}-2}[(A-1)(Z_{P}-1)\nonumber\\
~~~~~~~~~~~~~~~~~\times (G_{22}(b)-G^{2}_{00}(b))+(Z_{P}-1)\nonumber\\
~~~~~~~~~~~~~~~~~\times (G_{21}(b)-G^{2}_{00}(b))+(A-1)~~\nonumber\\
~~~~\times (G_{12}(b)-G^{2}_{00}(b))],
\label{eq34}
%\end{equation}
%\end{widetext}
\end{gather}
The quantities $G_{22}$, $G_{21}$, $G_{12}$, and $G_{00}$ in the above equation assume similar expressions as in Eq. (\ref{eq10}), but now they involve only the projectile proton density instead of projectile matter density.

\section{Results and Discussion}
\label{sec3}

Following the approach outlined in Sec. \ref{sec2}, we have analysed the charge-changing and interaction cross sections
for Z $\leq$ 9 isotopes on a $^{12}$\rm C target at a wider range of incident energies (200 - 1050 MeV/nucleon). The calculations involve different density distributions (SDHO, 2pF, 3pF) for projectile nuclei and use the in-medium as well as free behavior of the basic input of the Glauber model, the nucleon-nucleon (NN) amplitude. For the target $^{12}$\rm C nucleus, we involve the charge density as obtained from the electron scattering experiment \cite{21} and assume the neutron and proton densities to be the same. Moreover, at medium energies where the contribution due to inelastic scattering is negligible \cite{22}, the interaction cross section ($\sigma_{I}$) can be assumed to be nearly equal to the reaction cross section ($\sigma_{R}$). Hence the Glauber model S-matrix ($S_{el}(b)$) in Eq. (\ref{eq2}) can be used to calculate both $\sigma_{I}$ as well as $\sigma_{R}$ from Eq. (\ref{eq1}). 

 The $NN$ scattering amplitude that takes care of the nuclear in-medium effects, arising due to phase variation, higher momentum transfer components,
and Pauli blocking is parametrized as follows \cite{23}:
%\begin{widetext}
%\begin{equation}
\begin{gather}
f_{NN}(\vec q)=\biggl[\frac{ik\sigma_{NN}}{4\pi}\sum^{\infty}_{n=0}A_{n+1}\left(\frac{\sigma_{NN}}
{4\pi\beta_{NN}}\right) ^{n}\frac{(1-i\rho_{NN})^{n+1}}{(n+1)}\nonumber\\
~~~~~~~~~~~~\times \exp\left(\frac{-\beta_{NN}q^{2}}{2(n+1)}\right)\biggr]\exp\left(\frac{-i\gamma_{NN} q^{2}}{2}\right),
\label{eq17}
%\end{equation}
%\end{widetext}
\end{gather}
where
%\begin{widetext}
%\begin{equation}
\begin{gather}
A_{n+1}=\frac{A_{1}}{n(n+1)}+\frac{A_{2}}{(n-1)n}~~~~~~~~~~~~~~~~\nonumber\\
~~~~~~~~~+\frac{A_{3}}{(n-2)(n-1)}+.....+\frac{A_{n}}{1.2}~,
\label{eq18}
%\end{equation}
%\end{widetext}
\end{gather}
with $A_{1}=1$.\\

The $NN$ amplitude [Eq. (\ref{eq17})] consists of four parameters; 
$\sigma_{NN}$, $\rho_{NN}$, $\beta_{NN}$, and $\gamma_{NN}$. The values of these parameters at energies of our interest (700-1050) have been obtained by a linear interpolation and extrapolation of their values at 650, 800, and 1000 MeV \cite{12}. The first term in the NN amplitude [Eq. (\ref{eq17})], with n = 0 and $\gamma_{NN}$ = 0, gives the usually parametrized one-term NN amplitude \cite{24} that can be used to describe the behavior of the free NN amplitude:   
\begin{equation}
	f_{NN}(\vec q)=\frac{ik\sigma_{NN}}{4\pi}
	(1-i\rho_{NN})\exp\left(-\beta_{NN}q^{2}/2\right).
	\label{eq19}
	%\end{equation}
	%\end{widetext}
\end{equation}
The $NN$ amplitude [Eq. (\ref{eq19})] consists of three parameters; $\sigma_{NN}$, $\rho_{NN}$, and $\beta_{NN}$. The values of these parameters at the desired energies (200-1050 MeV) have been obtained by a linear interpolation and extrapolation of their values given in Ref. \cite{24}.
 
The intrinsic proton and neutron densities of the projectile, used in this work, have been taken in the following forms:

(i) SDHO density: The SDHO density has been obtained from the Slater determinant consisting of the harmonic oscillator single-particle wave functions \cite{12}. It involves the oscillator constant ($\rm \alpha^2$) as its basic input, which assumes different values for proton ($\rm \alpha_p^2$) and neutron ($\rm \alpha_n^2$) density distributions, giving rise to rms proton and neutron radii of a given nucleus.

(ii) Two-parameter Fermi (2pF) density: The functional form of two-parameter Fermi density distribution is given by
\begin{equation}
 \rho(r)=\frac{\rho_{0}}{1+e^{(r-R)/a}},
 \label{eq20}
\end{equation}
where $a$ is the diffuseness parameter and $R$ = $r_{0}Z^{1/3}$(or$N^{1/3}$) \cite{25} ($r_{0}$=1.2 fm) is the radius parameter. The central density $\rho_{0}$ is determined by the normalization to the number of protons ($Z$) or neutrons ($N$). It is clear that the only variable parameter in 2pF distribution, which is responsible to alter the proton and neutron radii, is the diffuseness parameter $a$.
 
(iii) Three-parameter Fermi (3pF) density: A common alternative to the 2pF density distribution is the three-parameter Fermi (3pF) distribution, given by 
\begin{equation}
\rho(r)=\frac{\rho_{0}[1+w(r^2/R^2)]}{1+e^{(r-R)/a}},
\label{eq21}
\end{equation}
This density introduces a central depression parameter $w$ which allows the central density to be depressed or raised, depending on the sign of $w$. Other quantities in Eq. (\ref{eq21}) have the same meaning as in Eq. (\ref{eq20}). In this form of the density distribution, the central depression parameter $w$ and the diffuseness parameter $a$ are used to determine the proton and neutron radii.

\begin{table*}
	\caption{$\sigma_{cc}^{p}$ provides the contribution to charge-changing cross section due to projectile protons on a $^{12}$\rm C target at energy E, involving the in-medium NN amplitude. The parameters of two-parameter Fermi (2pF) and three-parameter Fermi (3pF) density distributions (not shown) correspond to the experimental projectile charge radius 
		($\rm \langle r^2_{ch}\rangle^{1/2}$) \cite{26,27}. The correction parameter
		$\rm \epsilon \left (=\sigma_{cc}^{exp}/\sigma_{cc}^{p} \right )$ is the ratio of the experimental charge-changing cross section ($\rm \sigma_{cc}^{exp}$) 
		and $\rm \sigma_{cc}^{p}$. $\rm \epsilon_{avg}$ denotes the average value of $\rm \epsilon$ for the considered isotopes of a given element. SDHO predictions are taken from Ref. \cite {19}.}
	\renewcommand{\tabcolsep}{0.14mm}
	\renewcommand{\arraystretch}{1.2}
	\label{tab1}
	\begin{ruledtabular}
		\begin{tabular}{ccccccccccccc}
			\\
			Projectile& E/A(MeV) & $\rm \sigma_{cc}^{exp}$(mb)& \multicolumn{3}{c}{$\rm 2pF$} &\multicolumn{3}{c}{$\rm 3pF $}&\multicolumn{3}{c}{$\rm SDHO$\cite{19}}\\
			\cline{4-6} \cline{7-9} \cline{10-12}\\
%			&(MeV.)& (mb)& & & & & & &\\
						& &	& $\rm \sigma_{cc}^{p}$(mb)  &$\rm \epsilon$  &$\rm \epsilon_{avg}$  & $\rm \sigma_{cc}^{p}$(mb) &$\rm \epsilon $ &$\rm \epsilon_{avg}$& $\rm \sigma_{cc}^{p}$(mb)  &$\rm \epsilon$  & $\rm \epsilon_{avg}$  \\
%			&&&(mb)&&&(mb)&&&(mb)&&	\\
		\\
			\hline\\
			
			$\rm ^{7}Be$     & 772 &706$\pm$8\cite{8}   & 657.9     &1.073           &1.075 &657.8&1.073&1.075&658.2 &1.073&1.073         \\
			$\rm ^{9}Be$      &921 &682$\pm$30\cite{8}   &639.4   &1.067   & &639.1&1.067&&640.3&1.065&               \\
			$\rm ^{10}Be$    &946  &670$\pm$10\cite{8}   &617.2   &1.085          & &616.9&1.086&&618.1&1.084&                \\
			$\rm ^{11}Be$   &962 &681$\pm$3 \cite{8}    &635.6    &1.071          & &635.5&1.072&&636.7&1.069&                 \\
			$\rm ^{12}Be$    &925 &686$\pm$3 \cite{8}   &637.2    &1.077          & &637.1&1.077&&638.2&1.075&               \\
			\\
			$\rm ^{10}B$    &925 &685$\pm$14\cite{7}   &680.3     &1.007          &1.023 &680.1&1.007&1.023&680.2&1.007&1.023           \\
			$\rm ^{11}B$    &932 &702$\pm$6\cite{7}   &676.1      &1.038          & &675.9&1.039&&676.0&1.038&                 \\
			\\
			$\rm ^{12}C$    &937  &733$\pm$7\cite{9}   &732.4     &1.001          &0.992&732.1&1.001&0.992&731.4&1.002&0.993             \\
			$\rm ^{13}C$     &828  &726$\pm$7\cite{9}   &737.4    &0.985         &   &737.1&0.985&&736.0&0.986&                 \\
			$\rm ^{14}C$    &900 &731$\pm$7\cite{9}   &738.6      &0.989           & &738.3&0.990&&737.4&0.991                 \\
			\\
			$\rm ^{14}N$    &932    &793$\pm$9\cite{10}   &787.9   &1.006          &1.010&787.4&1.007&1.011&785.8&1.009&1.014         \\
			$\rm ^{15}N$     &776 &816$\pm$20\cite{10}   &804.8    &1.014         &   &803.7&1.015&&801.8&1.018&               \\
			\\
			$\rm ^{16}O$   &857  &848$\pm$4\cite{11}   &858.4     &0.988        &0.997&857.1&0.989&0.998&854.4&0.992&1.002           \\
			$\rm ^{18}O$    &872  &879$\pm$5\cite{11}   &874.1    &1.006        &   &872.8&1.007&&869.6&1.011&             \\
			\\
			$\rm ^{19}F$   &930  &1016$\pm$10\cite{5}   &934.9    &1.087        &1.087    &933.3&1.089&1.089&928.9&1.094&1.094       \\	
		\end{tabular}
	\end{ruledtabular}
\end{table*} 
 
As mentioned earlier, we, in this work, proposed to verify and improve utility of our recent findings \cite{19} on rms proton and neutron radii for  4 $\leq$ Z $ \leq$ 9 isotopes. In order to have clear understanding of the proposed work, we present our results in the following manner:

 \begin{figure}
 	\begin{center}
 		\includegraphics[height=15.0cm, width=8.9cm]{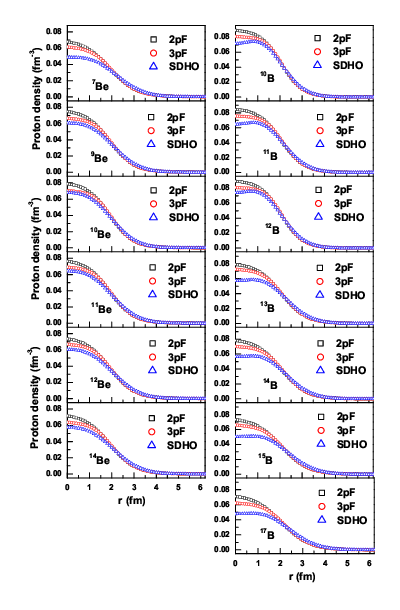}
 		\caption{SDHO, 2pF, and 3pF proton density distributions in $^{7,9-12,14}$\rm Be and $^{10-15,17}$\rm B isotopes. Squares, circles, and triangles show, respectively, 2pF, 3pF, and SDHO densities. The calculations with SDHO densities correspond to the values of oscillator parameters, taken from Ref. \cite{19}. The 2pF and 3pF density distributions correspond to proton radii given in Table~\ref{tab2}.}  
 		\label{fig1}
 	\end{center}
 \end{figure}

 \begin{figure}
 	\begin{center}
 	\includegraphics[height=16.5cm, width=8.6cm]{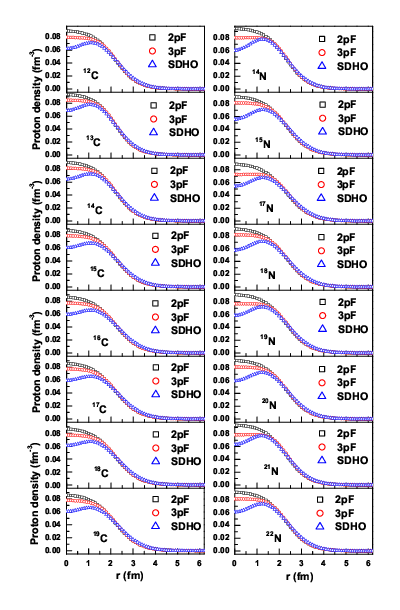}
 		\caption{Same as in Fig. 1, but for $^{12-19}$\rm C and $^{14,15,17-22}$\rm N isotopes.}  
 		\label{fig2}
 	\end{center}
 \end{figure}
 
 \begin{figure}
 	\begin{center}
 	\includegraphics[height=16.5cm, width=8.6cm]{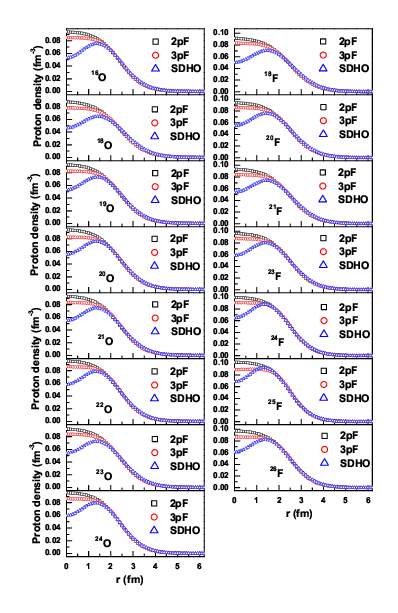}
 		\caption{Same as in Fig. 1, but for $^{16,18,19-24}$\rm O and $^{18,20,21,23-26}$\rm F isotopes.}  
 		\label{fig3}
 	\end{center}
 \end{figure}
 
 \begin{figure}
 	\begin{center}
 		\includegraphics[height=16.5cm, width=8.6cm]{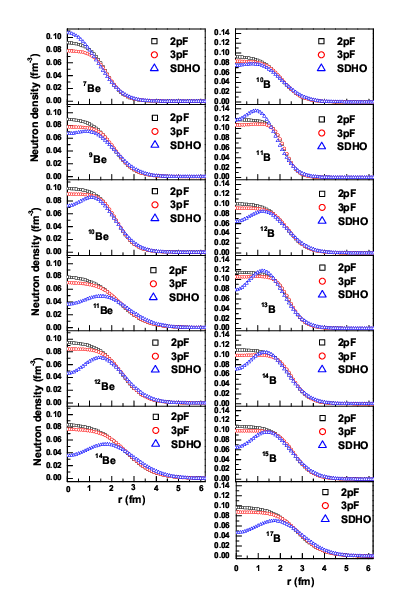}
 		\caption{SDHO, 2pF, and 3pF neutron density distributions in 
 			$^{7,9-12,14}$\rm Be and $^{10-15,17}$\rm B isotopes. Squares, circles, and triangles show, 
 			respectively, 2pF, 3pF, and SDHO densities. The calculations of SDHO densities correspond to the values of oscillator 
 			parameters, taken from Ref. \cite{19}. The 2pF and 3pF density distributions correspond to neutron radii given in Table~\ref{tab3}.}  
 		\label{fig4}
 	\end{center}
 \end{figure}

 \begin{figure}
 	\begin{center}
 		\includegraphics[height=16.5cm, width=8.6cm]{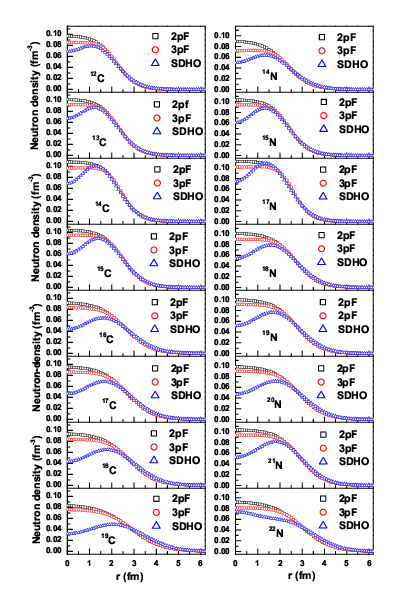}
 		\caption{Same as in Fig. 4, but for $^{12-19}$\rm C and $^{14,15,17-22}$\rm N isotopes.}  
 		\label{fig5}
 	\end{center}
 \end{figure}
 
\begin{figure}
	\begin{center}
		\includegraphics[height=16.5cm, width=8.6cm]{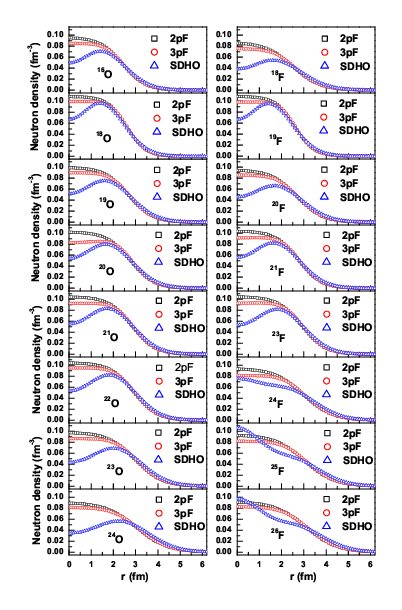}
		\caption{Same as in Fig. 4, but for $^{16,18,19-24}$\rm O and $^{18-21,23-26}$\rm F isotopes.}  
		\label{fig6}
	\end{center}
\end{figure} 

\subsection{Dependence of charge-changing and interaction cross sections on projectile density distributions, involving in-medium NN amplitude: Energy range 740-1050 MeV/nucleon}

To start with, let us make it clear that throughout the present work, the calculations have not involved any experimental uncertainty in predicting the proton(neutron) radius and charge-changing(interaction) cross section. Therefore, our theoretical results will be presented without any uncertainty. As a result, the predicted charge-changing and interaction cross sections may be better judged with reference to the central values of the corresponding experimental data. 

As mentioned in Sec.\ref{sec2}B, the calculation of charge-changing cross section ($\sigma_{cc}$) involves a phenomenological correction parameter $\epsilon$ to account for the contribution due to the presence of neutrons in the projectile. It has also been pointed out that due to some limitations in finding the correction parameter $\epsilon$ in some earlier works \cite{6,13,18}, we have adopted \cite{19} a different prescription to get the correction parameter.  In our approach, we first involve only those projectile nuclei for which the experimental charge radii \cite{26,27} as well as their experimental charge-changing cross sections are known. For such nuclei, we adjust the parameters of the considered densities to get their proton densities that lead to the experimentally known charge radii. These proton distributions are, then, used to compute the contribution to CCCS due to the projectile protons, $\sigma^{p}_{cc}$, using Eqs. (\ref{eq31})-(\ref{eq34}), and find the value of $\epsilon$ (=$\sigma^{exp}_{cc}/\sigma^{p}_{cc}$) in each case. The results of such calculations with a $^{12}$\rm C target using 2pF and 3pF density distributions are presented in Table \ref{tab1}. Table \ref{tab1} also presents the corresponding SDHO results \cite{19} for comparison. It is found that the correction parameter $\epsilon$ within the isotopic chain of a given element using 2pF and 3pF distributions follows similar trend as observed in the case of SDHO results. Interestingly, the average values of the correction parameter, $\epsilon_{avg}$, for 2pF and 3pF densities come closer to that obtained using SDHO density \cite{19} for each element. The last result is expected to provide proton density independence of the rms proton radii predicted in the analysis of charge-changing cross sections. To ascertain this result, following Ref. \cite{19}, we extract the rms proton radii by adjusting the parameters of 2pF and 3pF proton density distributions, involving the average value of the correction factor (Table \ref{tab1}) and the in-medium NN amplitude, that fit the experimental charge-changing cross sections. The values of the predicted rms proton radii ($r_{p}$) along with the corresponding proton density parameters are given in Table \ref{tab2}; $r_{p}$ takes  care of the finite size of the nucleon, giving rise to predicted charge radii for all the isotopes. Table \ref{tab2} also presents the corresponding results obtained using SDHO density \cite{19}. Comparison of the proton radii obtained by the 2pF and 3pF densities with those by the SDHO shows that the difference is well within 1$\%$. This small deviation in proton radii clearly suggests that, in our analysis of CCCSs, the use of different projectile (proton) density distributions are able to provide equivalent estimates for rms proton radii of all the isotopes under consideration. Moreover, this result provides confidence in using the rms proton radii as predicted in our recent work \cite{19} using the SDHO density. To make our results more transparent, we present, in Figs. \ref{fig1}-\ref{fig3}, our 2pF and 3pF proton density distributions along with the SDHO one \cite{19}, obtained using the respective density parameters given in Table \ref{tab2}. It is noticed that except for some deviations in the nuclear interior region, the SDHO, 2pF, and 3pF proton density distributions provide nearly a similar description of the surface part. This not only indicates but also strengthens our expectations that nucleus-nucleus collision mainly explores the surface region of the nucleus even at fairly large energies, thus suggesting that, in the present approach of nucleus-nucleus CCCS calculation, the smaller deviation in proton radii obtained using different proton density distributions is an indication of having the proton densities lying close to each other in the surface region. 

Next, we proceed to extract the rms neutron radii from the analysis of interaction cross sections, involving different choices for the neutron density distribution of the projectile; our aim, in this study, is to test the reliability of our rms neutron radii obtained using the SDHO density  \cite{19}. Keeping the similar values of the parameters of 2pF and 3pF proton densities (Table \ref{tab2}) and involving the in-medium NN amplitude, we now extract the rms neutron radii by adjusting the parameters of 2pF and 3pF densities for neutron distribution, that fit the corresponding experimental interaction cross sections. The values of the predicted rms neutron radii ($r_{n}$) along with the corresponding neutron density parameters are given in Table \ref{tab3}; Table \ref{tab3} also presents the corresponding results obtained using SDHO density \cite{19} for comparison, and it should be noted that $r_{n}$ takes  care of the finite size of the nucleon. Comparison of 2pF and 3pF predicted neutron radii with those obtained using SDHO density \cite{19} shows that, unlike the proton case, the difference in neutron radii appears to be a bit larger, within 3$\%$. To understand this relatively larger difference in neutron radii, we present, in Figs. \ref{fig4}-\ref{fig6}, the SDHO, 2pF, and 3pF neutron density distributions, obtained using the respective density parameters given in Table \ref{tab3}. It is seen that the deviation between the behavior of 2pF(3pF) and SDHO neutron distributions in the surface region is not as similar as it was noticed in the proton case, thereby expecting a bit larger difference between the 2pF(3pF) and SDHO predicted neutron radii (Table \ref{tab3}) as compared to that obtained in the proton case (Table \ref{tab2}). To look into the probable cause of having a relatively larger difference between the SDHO and 2pF(3pF) neutron radii as compared to proton, we first note that the deviation in proton radii, encountered in the analysis of CCCSs (Table \ref{tab2}), may be carried away by the neutron radius, which is treated as an adjustable parameter through the variation of the corresponding neutron density parameters, to fit the experimental interaction cross sections. As a result, it might happen that this interplay between the proton and neutron radii may contribute, to some extent, to already existing deviation in neutron radii; obviously, such an argument is not enough to explain the deviation up to $\sim$3$\%$ in neutron radii between the SDHO and 2pF(3pF) density distributions. Thus, it becomes necessary to explore more explanations of deviation and its possible origins so that one can have conclusive understanding of the SDHO and 2pF(3pF) results. In this connection, we notice that the deviation in neutron radii becomes larger in halo nuclei such as $^{11}$\rm Be, $^{14}$\rm Be, and $^{19}$\rm C, and neutron-rich C, N, O, and F isotopes with skins. Moreover, it is also observed that the neutron radii in such cases obtained with the SDHO are larger than those with the 2pF(3pF) distribution. This deviation can be related to the larger density distributions for the SDHO compared to the 2pF(3pF) model at large radial distances, as seen in Figs. \ref{fig4}-\ref{fig6}. Here, it can be added that the larger density distributions for SDHO at large radial distances come from smaller values of the oscillator parameter $\rm \alpha_n^2$ for neutron-rich nuclei obtained from the analysis of experimental interaction cross sections compared to the standard values of the oscillator parameter for normal nuclei \cite{28}. Considering these findings, the deviation in neutron radii can be attributed to the difference in the radial distributions between SDHO and 2pF(3pF) at large radial distances along with the use of small values of oscillator parameter $\rm \alpha_n^2$ in the SDHO distribution determined from the experimental interaction cross sections. However, we still hope that if the nucleus-nucleus collision probes nearly similar neutron distributions at large radial distances, the difference in neutron radii between different density distributions could be minimized. In the present context, it may, however, be added that if we compromise within 3$\%$ difference in the neutron radii between the 2pF(3pF) and SDHO densities, one may still believe that the SDHO neutron radii \cite{19} could be used for further studies.  

\begin{table*}
	\caption{Projectile (proton) density distribution parameters obtained from the fitting of experimental charge-changing cross sections, $\rm \sigma_{cc}^{exp}$, on a $^{12}$\rm C target to 2pF and 3pF at energy E using the in-medium NN amplitude and average value of the correction parameter $\rm \epsilon_{avg}$ (Table \ref{tab1}); $r_{p}$ gives the rms proton radius. The last column presents the corresponding results obtained using SDHO density \cite {19}.}
	\renewcommand{\tabcolsep}{0.14mm}
	\renewcommand{\arraystretch}{1.2}
	\label{tab2}
	\begin{ruledtabular}
		\begin{tabular}{ccccccccccccc}
	%		\\
			Projectile& E/A(MeV) & $\rm \sigma_{cc}^{exp}$(mb)& \multicolumn{2}{c}{$\rm 2pF$} &\multicolumn{3}{c}{$\rm 3pF $}&\multicolumn{2}{c}{$\rm SDHO$\cite{19}}\\
			\cline{4-5} \cline{6-8} \cline{9-10}
	%			&(MeV.)& (mb)& & & & & & &\\
			& &	& $a$(fm)  &$r_{p}$(fm)  & $a$(fm) &$w $ &$r_{p}$(fm)& $\rm \alpha_p^2(fm^{-2})$ &$r_{p}$(fm)   \\
			%			&&&(mb)&&&(mb)&&&(mb)&&	\\
	%		\\
			\hline
		\\
		$\rm ^{7}Be$   &772&706$\pm$8\cite{8}&0.589&2.6402&0.530&0.141  &2.6402 &0.2554  &2.6444                                       \\
		$\rm ^{9}Be$   &921&682$\pm$30\cite{8}&0.539&2.4871&0.484&0.163  &2.4871 &0.2965  &2.4865                          \\
		$\rm ^{10}Be$  &946&670$\pm$10\cite{8}&0.511 &2.4037&0.461&0.157 &2.4037 &0.3205  &2.4025                            \\
		$\rm ^{11}Be$  &962&681$\pm$3\cite{8}&0.528 &2.4543&0.480&0.138 &2.4543  &0.3097 &2.4530                            \\
		$\rm ^{12}Be$  &925&686$\pm$3\cite{8}&0.547&2.5107 &0.497&0.133  &2.5107 &0.2978  &2.5094                            \\
		$\rm ^{14}Be$  &833&697$\pm$4\cite{8}&0.562&2.5576 &0.504&0.160  &2.5576 &0.2886  &2.5612                            \\
		\\
		$\rm ^{10}B$   &925&685$\pm$14\cite{7}&0.472&2.3668&0.434&0.139  &2.3668 &0.3470  &2.3704                              \\
		$\rm ^{11}B$   &932&702$\pm$6\cite{7}&0.507&2.4655 &0.461&0.156 &2.4655  &0.3230 &2.4655                            \\
		$\rm ^{12}B$   &991&691$\pm$13\cite{7}&0.472&2.3663&0.434&0.139  &2.3663 &0.3529  &2.3657                           \\
		$\rm ^{13}B$   &897&723$\pm$6\cite{7}&0.548&2.5834 &0.504&0.120 &2.5829  &0.2965 &2.5872                           \\
		$\rm ^{14}B$   &926&727$\pm$4\cite{7}&0.555&2.6050 &0.504&0.148 &2.6050  &0.2930 &2.6080                           \\
		$\rm ^{15}B$   &920&747$\pm$5\cite{7}&0.593&2.7176 &0.536&0.146 &2.7171  &0.2707 &2.7181                            \\
		$\rm ^{17}B$   &862&759$\pm$4\cite{7}&0.610&2.7701 &0.541&0.182 &2.7701  &0.2618 &2.7721                          \\
		\\
		$\rm ^{12}C$    &937&733$\pm$7\cite{9}&0.499&2.5071 &0.452&0.169 &2.5060 &0.3255  &2.5045                         \\
		$\rm ^{13}C$    &828&726$\pm$7\cite{9}&0.471&2.4341&0.435&0.141 &2.4351  &0.3455 &2.4367                           \\
		$\rm ^{14}C$    &900&731$\pm$7\cite{9}&0.494&2.4956 &0.454&0.145 &2.4965 &0.3306  &2.4961                           \\
		$\rm ^{15}C$    &907&743$\pm$7\cite{9}&0.517&2.5582 &0.476&0.136 &2.5592 &0.3156  &2.5589                            \\
		$\rm ^{16}C$    &907&748$\pm$7\cite{9}&0.526&2.5822 &0.479&0.156 &2.5822 &0.3109  &2.5821                            \\
		$\rm ^{17}C$    &979&754$\pm$7\cite{9}&0.528&2.5899 &0.484&0.145 &2.5906 &0.3101  &2.5889                           \\
		$\rm ^{18}C$    &895&747$\pm$7\cite{9}&0.523&2.5756 &0.475&0.163 &2.5756 &0.3143  &2.5746                           \\
		$\rm ^{19}C$    &895&749$\pm$9\cite{9}&0.526&2.5836 &0.483&0.141 &2.5843 &0.3128  &2.5833                            \\
		\\
		$\rm ^{14}N$    &932&793$\pm$9\cite{10}&0.490&2.5454 &0.435&0.244 &2.5458 &0.3260  &2.5424                        \\
		$\rm ^{15}N$    &776&816$\pm$20\cite{10}&0.518&2.6196&0.474&0.154  &2.6196&0.3074   &2.6227                            \\
		$\rm ^{17}N$    &938&819$\pm$5\cite{10}&0.534&2.6656 &0.465&0.294 &2.6672 &0.2990  &2.6664                            \\
		$\rm ^{18}N$    &927&810$\pm$6\cite{10}&0.519&2.6234 &0.476&0.149 &2.6234 &0.3095  &2.6238                            \\
		$\rm ^{19}N$    &896&809$\pm$5\cite{10}&0.516&2.6146 &0.456&0.249 &2.6154 &0.3124  &2.6143                            \\
		$\rm ^{20}N$    &891&808$\pm$5\cite{10}&0.514&2.6083 &0.467&0.169 &2.6083 &0.3145  &2.6080                            \\
		$\rm ^{21}N$    &876&799$\pm$7\cite{10}&0.497&2.5626 &0.438&0.260 &2.5638 &0.3268  &2.5606                           \\
		$\rm ^{22}N$    &851&810$\pm$7\cite{10}&0.510&2.5989 &0.464&0.169 &2.5989 &0.3177  &2.5993                            \\
		\\
		$\rm ^{16}O$    &857&848$\pm$4\cite{11}&0.514&2.6658 &0.473&0.145 &2.6652 &0.3036  &2.6651                            \\
		$\rm ^{18}O$    &872&879$\pm$5\cite{11}&0.565&2.8054 &0.513&0.164 &2.8034 &0.2749  &2.8077                            \\
		$\rm ^{19}O$    &956&852$\pm$7\cite{11}&0.527&2.6998 &0.480&0.164 &2.6987 &0.2983  &2.6977                           \\
		$\rm ^{20}O$    &880&846$\pm$4\cite{11}&0.516&2.6697 &0.471&0.158 &2.6677 &0.3056  &2.6677                            \\
		$\rm ^{21}O$    &937&847$\pm$6\cite{11}&0.521&2.6839 &0.476&0.159 &2.6815 &0.3033  &2.6800                            \\
		$\rm ^{22}O$    &937&837$\pm$3\cite{11}&0.505&2.6406 &0.465&0.142 &2.6386 &0.3143  &2.6346                            \\
		$\rm ^{23}O$    &871&857$\pm$8\cite{11}&0.530&2.7096 &0.489&0.132 &2.7076 &0.2980  &2.7076                            \\
		$\rm ^{24}O$    &866&839$\pm$11\cite{11}&0.501&2.6318 &0.460&0.150 &2.6292&0.3168   &2.6278                            \\
		\\
		$\rm ^{18}F$    &930&998$\pm$25\cite{5}&0.557&2.8314 &0.514&0.133 &2.8304 &0.2880  &2.8294                            \\
		$\rm ^{20}F$    &930&980$\pm$13\cite{5}&0.533&2.7689 &0.493&0.132 &2.7679 &0.3029  &2.7639                            \\
		$\rm ^{21}F$    &930&986$\pm$10\cite{5}&0.542&2.7911 &0.497&0.147 &2.7895 &0.2979  &2.7891                            \\
		$\rm ^{23}F$    &930&967$\pm$22\cite{5}&0.516&2.7237 &0.478&0.137 &2.7239 &0.3146  &2.7177                            \\
		$\rm ^{24}F$    &930&946$\pm$24\cite{5}&0.486&2.6468 &0.449&0.141 &2.6425 &0.3346  &2.6368                            \\
		$\rm ^{25}F$    &930&934$\pm$54\cite{5}&0.469&2.6034 &0.426&0.196 &2.6048 &0.3468  &2.5914                            \\
		$\rm ^{26}F$    &930&962$\pm$48\cite{5}&0.509&2.7042 &0.465&0.166 &2.7042 &0.3207  &2.6961                            \\
	\end{tabular}
	\end{ruledtabular}
	\end{table*} 

\begin{table*}
	\caption{Projectile (neutron) density distribution parameters obtained from the fitting of experimental interaction cross sections, $\rm \sigma_{I}^{exp}$, on a $^{12}$\rm C target to 2pF and 3pF at energy E using the in-medium NN amplitude; $r_{n}$ gives the rms neutron radius. The last column presents the corresponding results obtained using SDHO density \cite {19}. The values of $\rm \sigma_{I}^{exp}$, with superscript $a$, are taken from Ref. \cite{30}}
	\renewcommand{\tabcolsep}{0.14mm}
	\renewcommand{\arraystretch}{1.2}
	\label{tab3}
	\begin{ruledtabular}
		\begin{tabular}{ccccccccccccc}
			%		\\
			Projectile& E/A(MeV) & $\rm \sigma_{I}^{exp}$(mb)& \multicolumn{2}{c}{$\rm 2pF$} &\multicolumn{3}{c}{$\rm 3pF $}&\multicolumn{2}{c}{$\rm SDHO$\cite{19}}\\
			\cline{4-5} \cline{6-8} \cline{9-10}
			%			&(MeV.)& (mb)& & & & & & &\\
			& &\cite{29}	& $a$(fm)  &$r_{n}$(fm)  & $a$(fm) &$w $ &$r_{n}$(fm)& $\rm \alpha_n^2(fm^{-2})$ &$r_{n}$(fm)   \\
			%			&&&(mb)&&&(mb)&&&(mb)&&	\\
			%		\\
			\hline
			\\
		$\rm ^{7}Be$    &790&738$\pm$9&0.386&1.9621&0.377&0.143  &2.0322 &0.4050   &1.9994                                       \\
		$\rm ^{9}Be$   &790&806$\pm$9&0.469&2.3606 &0.426&0.196 &2.3725  &0.3340  &2.4059                          \\
		$\rm ^{10}Be$   &790&813$\pm$10&0.426&2.3159 &0.398&0.129&2.3189  &0.3655  &2.3489                            \\
		$\rm ^{11}Be$   &790&942$\pm$8&0.611&2.8839 &0.553&0.162 &2.8879   &0.2391  &2.9479                            \\
		$\rm ^{12}Be$  &790&927$\pm$18&0.511&2.6570 &0.473&0.149  &2.6684  &0.2879  &2.7168                            \\
		$\rm ^{14}Be$  &800&1082$\pm$34&0.644&3.1217  &0.592&0.126 &3.1147 &0.2346   &3.1937                            \\
		\\
		$\rm ^{10}B$   &960&789$\pm$16&0.451&2.3101 &0.418&0.144 &2.3181  &0.3564  &2.3391                              \\
		$\rm ^{11}B$   &950&778$\pm$30&0.291&2.0053 &0.275&0.139 &2.0153   &0.4989 &2.0174                            \\
		$\rm ^{12}B$   &790&866$\pm$7&0.442&2.4213 &0.415&0.128 &2.4302   &0.3446 &2.4623                           \\
		$\rm ^{13}B$   &790&883$\pm$14&0.349&2.2662 &0.329&0.125 &2.2719   &0.4045 &2.2972                           \\
		$\rm ^{14}B$   &790&929$\pm$26&0.400&2.4398 &0.374&0.158 &2.4508   &0.3726 &2.4746                           \\
		$\rm ^{15}B$   &740&965$\pm$15&0.437&2.5778 &0.411&0.123 &2.5844   &0.3475 &2.6280                            \\
		$\rm ^{17}B$   &800&1118$\pm$22&0.562&2.9831 &0.523&0.140 &2.9931  &0.2737  &3.0693                          \\
		\\
		$\rm ^{12}C$    &950&853$\pm$6&0.444&2.3617 &0.409&0.176 &2.3759  &0.3487  &2.4197                         \\
		$\rm ^{13}C$    &960&862$\pm$12&0.434&2.4007 &0.408&0.128 &2.4107  &0.3525  &2.4401                           \\
		$\rm ^{14}C$    &965&880$\pm$19&0.399&2.3795 &0.375&0.152  &2.3925  &0.3672  &2.4157                           \\
		$\rm ^{15}C$    &740&945$\pm$10&0.451&2.5609 &0.426&0.124  &2.5709   &0.3345 &2.6159                            \\
		$\rm ^{16}C$    &960&1036$\pm$11&0.574&2.9255  &0.535&0.129 &2.9355   &0.2666 &3.0043                            \\
		$\rm ^{17}C$    &965&1056$\pm$10&0.562&2.9400 &0.523&0.139  &2.9499   &0.2753 &3.0151                           \\
		$\rm ^{18}C$    &955&1104$\pm$15&0.593&3.0633  &0.548&0.151 &3.0732   &0.2608 &3.1473                           \\
		$\rm ^{19}C$    &960&1231$\pm$28&0.708&3.4201  &0.656&0.121 &3.4301   &0.2124 &3.5334                            \\
		\\
		$\rm ^{14}N$     &965&932$\pm$9&0.524&2.6367  &0.460&0.303  &2.6567  &0.2874  &2.7077                        \\
		$\rm ^{15}N$    &975&930$\pm$30&0.436&2.4668  &0.414&0.123  &2.4828   &0.3362 &2.5288                            \\
		$\rm ^{17}N$    &710&965$\pm$24&0.395&2.4826  &0.376&0.141 &2.5026   &0.3778 &2.5266                            \\
		$\rm ^{18}N$    &1020&1046$\pm$8&0.512&2.8103  &0.475&0.159 &2.8233   &0.3021 &2.8809                            \\
		$\rm ^{19}N$    &1005&1076$\pm$9&0.532&2.9040  &0.501&0.123 &2.9182   &0.2914 &2.9800                            \\
		$\rm ^{20}N$    &950&1121$\pm$17&0.565&3.0318  &0.532&0.118 &3.0418   &0.2732 &3.1178                            \\
		$\rm ^{21}N$    &1005&1114$\pm$9&0.526&2.9737  &0.495&0.137 &2.9902   &0.2926 &3.0457                           \\
		$\rm ^{22}N$    &965&1245$\pm$49&0.654&3.3423  &0.598&0.172 &3.3523   &0.2346 &3.4332                            \\
		\\
		$\rm ^{16}O$    &970&982$\pm$6&0.506&2.6439 &0.471&0.148  &2.6589   &0.2898 &2.7277                            \\
		$\rm ^{18}O$    &1050&1032$\pm$26&0.426&2.5523 &0.407&0.119  &2.5706   &0.3499 &2.6281                            \\
		$\rm ^{19}O$    &970&1066$\pm$9&0.527&2.8477  &0.495&0.129 &2.8617   &0.2927 &2.9295                           \\
		$\rm ^{20}O$    &950&1078$\pm$10&0.517&2.8674  &0.461&0.303 &2.8911  &0.2996  &2.9412                            \\
		$\rm ^{21}O$    &980&1098$\pm$11&0.509&2.8921  &0.475&0.165 &2.9121   &0.3015 &2.9699                            \\
		$\rm ^{22}O$    &965&1123$\pm$24$^{a}$&0.522&2.9625 &0.494&0.123  &2.9789  &0.2940  &3.0403                            \\
		$\rm ^{23}O$    &960&1216$\pm$41$^{a}$&0.602&3.2022 &0.556&0.160  &3.2162   &0.2557 &3.2900                            \\
		$\rm ^{24}O$    &965&1318$\pm$52&0.696&3.4895  &0.647&0.125 &3.4995  &0.2193  &3.5813                            \\
		\\
		$\rm ^{18}F$    &975&1100$\pm$50&0.609&2.9777  &0.557&0.164 &2.9927  &0.2412  &3.0917                             \\
		$\rm ^{19}F$    &985&1043$\pm$24&0.427&2.5554  &0.407&0.136 &2.5784   &0.3476 &2.6391                            \\
		$\rm ^{20}F$    &950&1113$\pm$11&0.569&2.9595  &0.534&0.125 &2.9745   &0.2686 &3.0605                            \\
		$\rm ^{21}F$    &1000&1099$\pm$12&0.501&2.8282 &0.467&0.167  &2.8474  &0.3056  &2.9142                            \\
		$\rm ^{23}F$    &1020&1148$\pm$16&0.521&2.9602 &0.484&0.163  &2.9752   &0.2928 &3.0482                            \\
		$\rm ^{24}F$    &1005&1253$\pm$23&0.643&3.3115 &0.587&0.186  &3.3287   &0.2385 &3.4085                            \\
		$\rm ^{25}F$    &1010&1298$\pm$31&0.665&3.4064  &0.612&0.164 &3.4224   &0.2304 &3.4954                            \\
		$\rm ^{26}F$    &950&1353$\pm$54&0.693 &3.5141  &0.647&0.125 &3.5301   &0.2174 &3.6233                            \\
		\end{tabular}
	\end{ruledtabular}
\end{table*} 

\subsection{Understanding the role of nuclear medium effects on charge-changing and interaction cross sections: Energy range 740-1050 MeV/nucleon}

To assess the role of nuclear medium effects, we now perform parameter free calculations for charge-changing and interaction cross sections using SDHO, 2pF, and 3pF density distributions, involving the free NN amplitude. For this, we use the same parameters of SDHO, 2pF and 3pF proton and neutron densities as given in Tables \ref{tab2} and Table \ref{tab3}.

\begin{figure}
	\begin{center}
		\includegraphics[height=10.6cm, width=8.6cm]{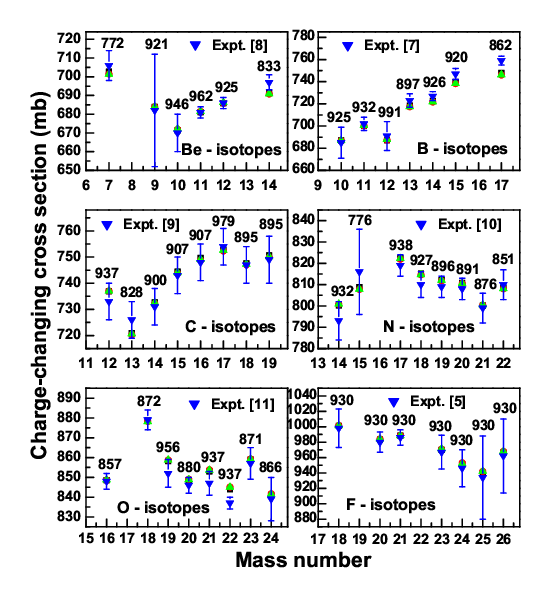}
		\caption{Charge-changing cross sections using the parameters of free NN amplitude \cite{24}. Filled squares, circles, and triangles correspond, 
			respectively, to SDHO, 2pF, and 3pF densities. The calculations with SDHO proton densities correspond to the values of oscillator 
			parameters taken from Ref. \cite{19}, and the calculations with 2pF and 3pF proton density distributions correspond to proton radii given in Table~\ref{tab2}. The number above each experimental data (shown by filled inverted triangles) represents the projectile energy/nucleon (in MeV).}
		\label{fig7}
	\end{center}
\end{figure}

For CCCS calculations, we first deduce the correction parameter as discussed above. The values of the correction parameter with the free NN amplitude are presented in Table \ref{tab4}. It is found that the features of the correction parameter are same as found in the results presented in Table \ref{tab1}. However, we noticed that the calculations for $\sigma^{p}_{cc}$ in Tables \ref{tab1} and \ref{tab4} though involve similar parameters for SDHO, 2pF, and 3pF densities, the average value of the correction parameter $\epsilon_{avg}$ for each element is found to be different for in-medium and free NN amplitudes. An obvious reason for getting different values of $\epsilon_{avg}$ for a given element is the difference between in-medium and free NN total cross sections; the in-medium NN total cross section is found to be less than the free one \cite{31}. As a result, the $\sigma^{p}_{cc}$, which is directly connected to the pp and pn total cross sections through NN scattering amplitude, has a larger value using the free NN amplitude as compared to the in-medium one, giving rise to different values of $\epsilon_{avg}$ for in-medium and free NN amplitudes. However, it comes out that the use of $\epsilon_{avg}$ corresponding to the free NN amplitude (Table \ref{tab4}) predicts CCCSs lying very close to the experimental data (Fig. \ref{fig7}). This result suggests that any change in $\sigma^{p}_{cc}$ due to change in the behavior of NN amplitude is being accommodated by the corresponding value of $\epsilon_{avg}$. In other words, we find that the use of SDHO, 2pF, and 3pF densities, involving our proton radii (\cite{19}, Table \ref{tab2}), provide equivalent results for CCCSs, irrespective of whether we use the in-medium or free NN amplitude. Thus, we observe that (i) our calculations for CCCSs are unable to distinguish between the results obtained using the in-medium and free NN amplitudes, suggesting that the importance of nuclear medium effects can not be assessed from the analysis of CCCSs, and (ii) the combined results of Sec. \ref{sec3}A and Sec. \ref{sec3}B on proton radii led us to believe that the proton radii, as obtained in Ref. \cite{19}, may be used in any realistic nuclear calculations.

\begin{figure}
	\begin{center}
		\includegraphics[height=10.6cm, width=8.6cm]{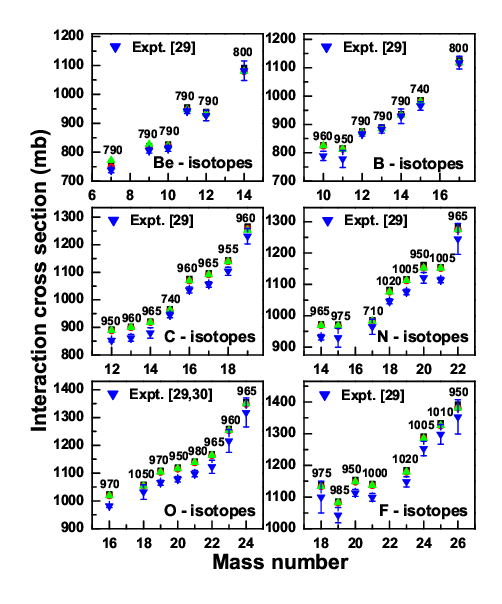}
		\caption{Interaction cross sections using the parameters of free NN amplitude \cite{24}. Filled squares, circles, and triangles correspond, 
			respectively, to SDHO, 2pF, and 3pF densities. The calculations with SDHO proton and neutron densities correspond to the values of oscillator 
			parameters taken from Ref. \cite{19}, and the calculations with 2pF and 3pF proton density distributions correspond to proton and neutron radii given in Tables \ref{tab2} and \ref{tab3}. The number above each experimental data (shown by filled inverted triangles) represents the projectile energy/nucleon (in MeV).}
		\label{fig8}
	\end{center}
\end{figure}

Our next goal is to assess how far the interaction cross section gets affected by the use of free NN amplitude instead of the in-medium one. Taking the parameters of SDHO, 2pF and 3pF proton and neutron distributions given in Tables \ref{tab2} and \ref{tab3}, we now predict the interaction cross sections using the free NN amplitude. The results are shown in Fig. \ref{fig8}. It is found that the predicted interaction cross sections show within 5$\%$ deviation from the experimental data. As mentioned in Sec. \ref{sec3}A, the parameters of SDHO, 2pF and 3pF proton and neutron distributions, given in Tables \ref{tab2} and \ref{tab3}, exactly reproduce the experimental interaction cross sections, using the in-medium NN amplitude. Therefore, loosely speaking, the observed deviation of 5$\%$ at most can be taken as a difference between the interaction cross sections obtained using the in-medium and free NN amplitudes, thereby suggesting the need of nuclear medium effects, entered through in-medium NN amplitude, in the analysis of interaction cross sections. Unfortunately, the in-medium NN amplitude is not available at the desired energies, whereas the parameters of the free NN amplitude are available in the literature at a wider range of energies. Due to this, the free NN amplitude is usually used in Glauber model calculations, and, in the present work, it also motivated us to use the free NN amplitude and see what could be said about the findings for charge-changing and interaction cross sections. In the light of this, we found that the CCCSs could be nicely explained using the free NN amplitude; the results also showed that CCCSs are insensitive to the choice of the proton density distribution. Whereas, in the case of interaction cross sections, the use of free NN amplitude though provides equivalent results for all the considered proton and neutron densities, but these results deviate within 5$\%$ from the experimental data.  The later result suggests that while using the free NN amplitude, along with our predicted parameters for SDHO, 2pF, and 3pF proton and neutron distributions (Tables \ref{tab2} and \ref{tab3}), one should keep in mind that the theoretical results for interaction cross sections may differ $\sim$5$\%$ at most from the experimental data. 

\subsection{Charge-changing and interaction cross sections at relatively lower energies} 

To move further, let us recall our calculations for CCCSs involving the free NN amplitude. These calculations demonstrated that the use of our predicted proton radii \cite{19} obtained with in-medium NN amplitude provide equivalent results for CCCSs when analysed with the free NN amplitude. Keeping this in view, we perform calculations for the CCCSs at relatively lower energies ($\sim$ 230, 300 MeV/nucleon) \cite{32,33} using the free NN amplitude and involving the SDHO proton radii (densities) obtained in Ref. \cite{19}. The aim for such a study is to see how well our predicted proton radii \cite{19} accommodate the CCCSs at lower energies and what could be said about the energy dependence of the correction parameter.

As already discussed, we start with finding out average value of the correction parameter $\epsilon_{avg}$ at $\sim$ 230 and 300 MeV/nucleon for C, N, and O isotopes using the recent experimental data \cite{32,33} on CCCSs. The results are presented in Table \ref{tab5}. These results along with those presented in Table \ref{tab4} clearly show the energy dependence of the correction parameter for the isotopes of a given element. Next, using the values of average correction parameter at $\sim$ 230 and 300 MeV/nucleon, we have performed calculations for CCCSs involving the free NN amplitude and considering the parameters of SDHO proton densities obtained in Ref. \cite{19}. The results are presented in Table \ref{tab6}.  It is found that the predicted CCCSs agree fairly well with the experiment. This result adds credibility for the use of SDHO proton radii \cite{19} in the study of charge-changing cross section at relatively lower incident energies. Moreover, we got motivated to predict (i) the rms proton radii of $^{11}$\rm C, $^{13,16}$\rm N, and $^{15,17}$\rm O, not obtained in our earlier work \cite{19}, and (ii) the CCCSs, at some random energies in the energy range 200-300 MeV/nucleon, for those isotopes whose experimental values are not available; these results are also given in Table \ref{tab6}, and we expect to have more experiments in future to verify our predictions on CCCSs.

Finally, we test our predicted proton and neutron radii (densities) \cite{19} in the analysis of reaction cross sections at lower energies compatible to the energies considered above in the calculations of CCCSs. Unfortunately, we do not have sufficient data on reaction cross sections for nuclear collisions with a $^{12}$\rm C target at lower energies, except for $^{12}$\rm C on $^{12}$\rm C at 200, 250, and 300 MeV/nucleon \cite{34}. The results are presented in Table \ref{tab7}, and are found to provide satisfactory explanation of the data within experimental errors.

\begin{table*}
	\caption{$\sigma_{cc}^{p}$ provides the contribution to charge-changing cross section due to projectile protons on a  $^{12}$\rm C target at energy E, involving the free NN amplitude. The parameters of SDHO, two-parameter Fermi (2pF) and three-parameter Fermi (3pF) density distributions (not shown) correspond to the experimental projectile charge radius 
		($\rm \langle r^2_{ch}\rangle^{1/2}$) \cite{26,27}. The correction parameter
		$\rm \epsilon \left (=\sigma_{cc}^{exp}/\sigma_{cc}^{p} \right )$ is the ratio of the experimental charge-changing cross section ($\rm \sigma_{cc}^{exp}$) 
		and $\rm \sigma_{cc}^{p}$. $\rm \epsilon_{avg}$ denotes the average value of $\rm \epsilon$ for the considered isotopes of a given element.}
	\renewcommand{\tabcolsep}{0.14mm}
	\renewcommand{\arraystretch}{1.2}
	\label{tab4}
	\begin{ruledtabular}
		\begin{tabular}{ccccccccccccc}
			\\ 
			Projectile& E/A(MeV) & $\rm \sigma_{cc}^{exp}$(mb)& \multicolumn{3}{c}{$\rm 2pF$} &\multicolumn{3}{c}{$\rm 3pF $}&\multicolumn{3}{c}{$\rm SDHO$\cite{19}}\\
			\cline{4-6} \cline{7-9} \cline{10-12}\\
			%			&(MeV.)& (mb)& & & & & & &\\
			& &	& $\rm \sigma_{cc}^{p}$(mb)  &$\rm \epsilon$  &$\rm \epsilon_{avg}$  & $\rm \sigma_{cc}^{p}$(mb) &$\rm \epsilon $ &$\rm \epsilon_{avg}$& $\rm \sigma_{cc}^{p}$(mb)  &$\rm \epsilon$  & $\rm \epsilon_{avg}$  \\
			%			&&&(mb)&&&(mb)&&&(mb)&&	\
			\\ 
			\hline\\
			
			$\rm ^{7}Be$     & 772 &706$\pm$8\cite{8}   & 673.6     &1.048         &1.043 &673.5&1.048&1.043&677.4 &1.042&1.038         \\
			$\rm ^{9}Be$      &921 &682$\pm$30\cite{8}   &660.1   &1.033          & &659.9&1.033&&663.3&1.028&               \\
			$\rm ^{10}Be$    &946  &670$\pm$10\cite{8}   &639.1   &1.048          & &638.9&1.048&&641.3&1.045&                \\
			$\rm ^{11}Be$   &962 &681$\pm$3 \cite{8}    &655.3    &1.039          & &655.3&1.039&&658.0&1.035&                 \\
			$\rm ^{12}Be$    &925 &686$\pm$3 \cite{8}   &656.9    &1.044          & &656.9&1.044&&660.0&1.039&               \\
			\\
			$\rm ^{10}B$    &925 &685$\pm$14\cite{7}   &704.8     &0.972          &0.987 &704.8&0.972&0.988&706.8&0.969&0.985           \\
			$\rm ^{11}B$    &932 &702$\pm$6\cite{7}   &699.8      &1.003          & &699.4&1.004&&701.6&1.001&                 \\
			\\
			$\rm ^{12}C$    &937  &733$\pm$7\cite{9}   &758.6     &0.966         &0.964 &758.3&0.967&0.964&759.7&0.965&0.962             \\
			$\rm ^{13}C$     &828  &726$\pm$7\cite{9}   &752.4    &0.965         &   &752.4&0.965&&753.8&0.963&                 \\
			$\rm ^{14}C$    &900 &731$\pm$7\cite{9}   &761.3     &0.960          & &761.2&0.960&&762.7&0.958                 \\
			\\
			$\rm ^{14}N$    &932    &793$\pm$9\cite{10}   &814.8   &0.973          &0.986&814.7&0.973&0.986&815.5&0.972&0.985        \\
			$\rm ^{15}N$     &776 &816$\pm$20\cite{10}   &816.9    &0.999         &   &816.8&0.999&&817.7&0.998&               \\
			\\
			$\rm ^{16}O$   &857  &848$\pm$4\cite{11}   &877.2     &0.967        &0.976&876.8&0.967&0.976&876.9&0.967&0.976           \\
			$\rm ^{18}O$    &872  &879$\pm$5\cite{11}   &892.9    &0.984        &   &892.1&0.985&&892.7&0.985&             \\
			\\
			$\rm ^{19}F$   &930  &1016$\pm$10\cite{5}   &958.6    &1.060        &1.060    &957.8&1.061&1.061&956.9&1.062&1.062       \\
		\end{tabular}
	\end{ruledtabular}
\end{table*}

\begin{table*}
	\caption{$\sigma_{cc}^{p}$ provides the contribution to charge-changing cross section due to projectile protons on a  $^{12}$\rm C target at energy E $\sim$ 230 and 300 MeV/nucleon \cite{32,33} , using the free NN amplitude and involving the SDHO density distribution. The parameter of SDHO (not shown) corresponds to the experimental projectile charge radius 
		($\langle r^2_{ch}\rangle^{1/2}$) \cite{26,27}. The correction parameter
		$\epsilon \left (=\sigma_{cc}^{exp}/\sigma_{cc}^{p} \right )$ is the ratio of the experimental charge-changing cross section ($\sigma_{cc}^{exp}$) 
		and $\sigma_{CC}^{p}$. $\rm \epsilon_{avg}$ denotes the average value of $\epsilon$ for the considered isotopes of a given element.}
	\label{tab5}
	\begin{tabular}{cccccc|cccccc}	
		\\
		\hline
		\hline
		&&&&&&&&&&&\\
		$\rm Projectile$  &$\rm E/A$      &$\rm \sigma_{cc}^{p}$  &$\rm  \sigma_{cc}^{exp}$ &$\rm \epsilon $                            & $\rm  \epsilon_{avg}$&	$\rm Projectile$  &$\rm E/A$      &$\rm \sigma_{cc}^{p}$  &$\rm  \sigma_{cc}^{exp}$ &$\rm \epsilon $                            & $\rm  \epsilon_{avg}$\\
		&$\rm (MeV)$    &$\rm (mb)$               &$\rm (mb)$               &                            &&&$\rm (MeV)$    &$\rm (mb)$               &$\rm (mb)$               &                            &\\
		&               &        &                &    &  &&&&&& \\        
		\hline
		&&&&&&&&&&&\\
		$E \sim230 MeV/nucleon$\cite{33}&&&&&&                            $E \sim300 MeV/nucleon$\cite{32}&&&&&\\  
		&&&&&&&&&&&\\                        
		$\rm ^{12}C$   &228            &704.3          &723$\pm$23                 &1.027                      &1.020&$\rm ^{12}C$   &294            &678.7          &731$\pm$52                 &1.080                      &1.079\\     
		$\rm ^{13}C$   &231            &700.3          &720$\pm$25                 &1.028                      &&$\rm ^{13}C$   &322            &674.1          &729$\pm$22                 &1.081                      &\\
		$\rm ^{14}C$   &234            &705.2          &707$\pm$13                 &1.003                      &&$\rm ^{14}C$   &339            &679.7          &732$\pm$22                 &1.077                      &\\
		&&&&&&&&&&&\\
		$\rm ^{14}N$   &223            &761.7          &843$\pm$32                 &1.106                      &1.079&$\rm ^{14}N$   &289            &733.5          &878$\pm$77                 &1.196                      &1.151\\     
		$\rm ^{15}N$   &226            &767.9          &808$\pm$15                 &1.052                      &&$\rm ^{15}N$   &315            &737.4          &815$\pm$11                 &1.105                      &\\
		&&&&&&&&&&&\\
		$\rm ^{16}O$   &219            &826.7          &862$\pm$17                 &1.043                      &1.043&$\rm ^{18}O$   &368            &801.8         &887$\pm$39                 &1.106                      &1.106\\
			\hline
			\hline
			\\
	\end{tabular}
\end{table*}

\begin{table*}[htb]
	\caption{Table shows our predicted charge-changing cross section, $\rm \sigma_{cc}^{pred}$, on a  $^{12}$\rm C target at energy E $\sim$ 230 and 300 MeV/nucleon \cite{32,33} using the (i) average value of the correction parameter $\rm \epsilon_{avg}$ (Table \ref{tab5}) and (ii) free NN amplitude, and involving the SDHO density distribution\cite{19}. We have also presented $\rm \sigma_{cc}^{pred}$ at some random energies, in the considered energy range, for those isotopes whose experimental charge-changing cross sections $\rm \sigma_{cc}^{exp}$ are not available. Superscript $a$ shows predicted proton radius $r_{p}$, obtained by reproducing $\rm \sigma_{cc}^{exp}$ using SDHO proton density distribution\cite{19}.}
	\centering
	\label{tab6}
		\begin{tabular}{cccc|ccccc}	
		\\
		\hline
		\hline
		&&&&&&&\\
		$\rm Projectile$  &$\rm E/A$      &$\rm \sigma_{cc}^{pred}$  &$\rm  \sigma_{cc}^{expt}$&$\rm Projectile$&$\rm E/A$&$\rm \sigma_{cc}^{pred}$&$\rm  \sigma_{cc}^{expt}$&$r_{p}$ \\
						  &$\rm (MeV)$    &$\rm (mb)$                &$\rm (mb)$& &$\rm (MeV)$&$\rm (mb)$&$\rm (mb)$&$\rm (fm)$               \\
																				               &                          &    & &&&&                     \\        
		\hline
	&&&&&&&\\
		~~~~~~~~~~~~~~		$E \sim230 MeV/nucleon$\cite{33}&&&&~~~~~~~~~~~~$E \sim300 MeV/nucleon$\cite{32}&&&\\
		               &&&&&&\\
		$\rm ^{12}C$   &228            &723.9 &723$\pm$23         &$\rm ^{11}C$ &319&716.0&716$\pm$20&2.3971$^{a}$\\
		&&&&&&&&2.32(11)\cite{33}\\	          	       
		$\rm ^{13}C$   &231            &709.4 &720$\pm$25         &$\rm ^{12}C$   &294&738.6&731$\pm$52                           \\     
		$\rm ^{14}C$   &234            &717.5 &707$\pm$13         &$\rm^{13}C$    &322&723.4&729$\pm$22                           \\
		$\rm ^{15}C$   &236            &727.8 &749$\pm$19         &$\rm^{14}C$    &339&732.6&732$\pm$22                           \\
		$\rm ^{16}C$   &237            &731.9 &738$\pm$17         &$\rm^{15}C$    &327&744.6&758$\pm$56                           \\     
	    $\rm ^{17}C$   &228            &739.6 &                   &$\rm^{16}C$    &300&750.8          &                           \\
		$\rm ^{18}C$   &228            &737.2          &          &$\rm^{17}C$    &300&751.7          &                \\
		$\rm ^{19}C$   &228            &739.7          &          &$\rm^{18}C$    &300&748.8          &                \\
		$\rm ^{14}N$   &223            &818.9 &843$\pm$32         &$\rm^{19}C$    &300&752.1         &-    \\
		$\rm ^{15}N$   &226            &832.5 &808$\pm$15         &$\rm ^{13}N$   &310&752.0&752$\pm$35&2.2856$^{a}$\\
		&&&&&&&&2.37(16)\cite{33}\\	
		$\rm ^{17}N$   &236            &833.8 &809$\pm$17         &$\rm ^{14}N$   &289&807.4&878$\pm$77               \\   
		$\rm ^{18}N$   &236            &826.1          &         &$\rm ^{15}N$   &315&818.2&815$\pm$11               \\       
		$\rm ^{19}N$   &236            &824.2          &          &$\rm ^{16}N$   &322&813.0&813$\pm$9&2.5995$^{a}$\\
		&&&&&&&&2.50(7)\cite{33}\\	
		$\rm ^{20}N$   &236            &823.1          &          &$\rm ^{17}N$   &328&825.2&790$\pm$11               \\
		$\rm ^{21}N$   &236            &812.9          &          &$\rm ^{18}N$   &300&818.8          &    \\        
		$\rm ^{22}N$   &236            &821.9          &          &$\rm ^{19}N$   &300&817.5          &     \\ 
		$\rm ^{16}O$   &219            &854.5         &862$\pm$17 &$\rm ^{20}N$   &300&816.4          &    \\             
		$\rm ^{18}O$   &219            &886.2         &           &$\rm ^{21}N$   &300            &806.4          &   \\           	
		$\rm ^{19}O$   &219            &861.9         &           &$\rm ^{22}N$   &300            &815.4          &     \\
	    $\rm ^{20}O$   &219            &855.1         &           &$\rm ^{15}O$   &301&880.0&880$\pm$18&2.7237$^{a}$\\ 
	    &&&&&&&&2.69(9)\cite{33}\\	
	    $\rm ^{21}O$   &219            &857.8         &           &$\rm ^{16}O$   &300&867.1         &              \\ 
	    $\rm ^{22}O$   &219            &848.3         &           &$\rm ^{17}O$   &308&866.0&866$\pm$11&2.6631$^{a}$\\ 
	    &&&&&&&&2.62(8)\cite{33}\\	
	    $\rm ^{23}O$   &219            &865.4         &           &$\rm ^{18}O$   &368&894.7         &887$\pm$39   \\                  
	    $\rm ^{24}O$   &219            &847.6          &          &$\rm ^{19}O$   &300&874.1         &     \\
	                   &               &               &          &$\rm ^{20}O$   &300&867.2         &    \\ 
	                   &               &               &          &$\rm ^{21}O$   &300&869.8         &    \\
		               &               &               &          &$\rm ^{22}O$   &300&860.5         &   \\
		               &               &               &          &$\rm ^{23}O$   &300&877.8         &  \\
	                   &               &               &          &$\rm ^{24}O$   &300&859.9         &    \\          
			\hline
			\hline
			\\	
	\end{tabular}
		\end{table*}
	
	\begin{table}
		\caption{Predicted reaction cross section, $\sigma_{R}^{pred}$, for $^{12}$\rm C on a $^{12}$\rm C target at energy E using the free NN amplitude and involving the SDHO proton and neutron density distributions \cite{19}. The experimental reaction cross section is shown by $\rm \sigma_{R}^{exp}$.}
		\label{tab7}
		\begin{tabular}{ccccccccc}		
			\\
			\hline
			\hline
			\\
			$\rm E/A$      &$\rm \sigma_{R}^{pred}$  &$\rm  \sigma_{R}^{exp}$\cite{34} \\
			$\rm (MeV)$    &$\rm (mb)$               &$\rm (mb)$                             \\
			&&\\      
			\hline
			\\ 	
			$200$           &855.4           &864$\pm$45                             \\     
			$250$           &809.0           &873$\pm$60                             \\
			$300$           &809.7           &858$\pm$60                             \\
			\hline
			\hline
			\\	
		\end{tabular}
	\end{table}
	
\section{Summary and Conclusions}
\label{sec4}

To improve utility of our recent findings \cite{19}, we, in this work, proposed to establish credibility for the use of Slater determinant harmonic oscillator (SDHO) density by analysing the charge-changing cross section ($\sigma_{cc}$) and interaction cross section ($\sigma_{I}$) for Be, B, C, N, O, and F isotopes on $^{12}$\rm C involving different density distributions, at a wider range of incident energies (200-1050 MeV/nucleon). The calculations also assess the role of nuclear medium effects. Working within the framework of Glauber model, we have first considered two-parameter Fermi (2pF) and three-parameter Fermi (3pF) shapes of density distributions to study $\sigma_{cc}$ and $\sigma_{I}$ in the energy range 740-1050 MeV/nucleon, involving the in-medium nucleon-nucleon (NN) amplitude. Following the scaling (correction) parameter approach \cite{19} for calculating the charge-changing cross section, we find that the predicted rms proton and neutron radii, obtained from the analysis of charge-changing cross section and interaction cross section using 2pF and 3pF densities, are in close agreement with those predicted by SDHO density distribution \cite{19}. This result strengthens our SDHO-based findings \cite{19} on rms proton and neutron radii in all the cases (Be-F). Next, we have investigated to see how far the use of free NN amplitude affects our findings obtained using the in-medium NN amplitude. Our calculations show that the correction parameter takes care of the difference between the free and in-medium NN amplitudes, giving rise to equivalent results for charge-changing cross section with both the forms of NN amplitude. This result shows that the importance of nuclear medium effects can not be assessed from the analysis of charge-changing cross sections. In the case of interaction cross sections, we find that the use of free NN amplitude overestimates the experimental data within 5$\%$. It was, therefore, concluded that one may not use the in-medium and free NN amplitudes on equal footings as far as the analysis of interaction cross section is concerned. This result highlights the importance of nuclear medium effects in the study of interaction cross sections.  Overall, the present results support our recent findings \cite{19} on proton and neutron radii obtained using SDHO densities.  

The successful use of free NN amplitude in calculating charge-changing cross section (CCCS) has motivated us to calculate CCCS at relatively lower energies, with the aim to test the reliability of proton radii obtained in our recent work \cite{19} and also to see the energy dependence of the correction parameter needed in the study of CCCS. For this, we have taken the available experimental data on CCCS for C, N, and O isotopes on $^{12}$\rm C at $\sim$ 230 and 300 MeV/nucleon \cite{32,33}. The results of such calculations are found to agree fairly well with the experimental data, and the correction parameter is found to be energy dependent for the isotopes of a given element. Finally, we focus our attention to test our predicted proton and neutron radii (densities) \cite{19} in the analysis of reaction cross section at lower energies. Unfortunately, we do not have sufficient data on reaction cross sections for nuclear collisions with a $^{12}$\rm C target at lower energies compatible to the energies considered in the analysis of CCCSs, except for $^{12}$\rm C on $^{12}$\rm C at 200, 250, and 300 MeV/nucleon \cite{34}. The results are presented in Table \ref{tab7}, and are found to provide satisfactory explanation of the data within experimental errors. Overall, the present work supports the SDHO predictions for proton
and neutron radii of Be, B, C, N, O, and F isotopes, as obtained in our recent work \cite{19}. Still, there is a need to include more data on higher atomic number (exotic) nuclei and predict their matter radii using SDHO densities.

\section{ Acknowledgments}

A.A.U. acknowledges the Inter-University Centre for Astronomy and Astrophysics, Pune, India for support via an associateship and for hospitality. Z.H. acknowledges the UGC-BSR Research Start-Up-Grant (No.F.30-310/2016(BSR)).

\end{document}